\documentclass[fleqn,10pt]{wlscirep}
\usepackage{multirow}
\usepackage[ruled,vlined,linesnumbered]{algorithm2e}
\usepackage{algorithmic}
\usepackage{makecell}

\title{Temporal Topic Modeling to Assess Associations between News Trends and Infectious Disease Outbreaks}

\author[1,*]{Saurav Ghosh}
\author[1]{Prithwish Chakraborty}
\author[2, 3, 4]{Elaine O. Nsoesie}
\author[2]{Emily Cohn}
\author[3, 6]{Sumiko R. Mekaru}
\author[2, 3, 5, 6]{John S. Brownstein}
\author[1]{Naren Ramakrishnan}

\affil[1]{Department of Computer Science, Virginia Tech, Arlington, Virginia, USA,}
\affil[2]{Children's Hospital Informatics Program, Boston Children’s Hospital, Boston, Massachusetts, USA,}
\affil[3]{Department of Pediatrics, Harvard Medical School, Boston, Massachusetts, USA,}
\affil[4]{Institute for Health Metrics and Evaluation, University of Washington, Seattle, Washington, USA,}
\affil[5]{Department of Epidemiology, Biostatistics and Occupational Health, McGill University, Montreal, Canada,}
\affil[6]{Epidemico, Inc., Boston, Massachsuetts, USA.}

\affil[*]{sauravcsvt@vt.edu}

\newcommand{\fullmodel}{\sf EpiNews}        
\newcommand{\fullmodelforecast}{\sf EpiNews-ARNet}
\newcommand{\fullmodelbaseline}{\sf EpiNews-ARMAX}                            
\newcommand{\baselinemodel}{\sf Casecount-ARMA}  


\begin{abstract}
In retrospective assessments, internet news reports have been shown to capture early reports of \textit{unknown} infectious disease transmission prior to official laboratory confirmation. In general, media interest and reporting peaks and wanes during the course of an outbreak. In this study, we quantify the extent to which media interest during infectious disease outbreaks is indicative of trends of reported incidence. We introduce an approach that uses supervised temporal topic models to transform large corpora of news articles into temporal topic trends. The key advantages of this approach include, applicability to a wide range of diseases, and ability to capture disease dynamics - including seasonality, abrupt peaks and troughs. We evaluated the method using data from multiple infectious disease outbreaks reported in the United States of America (U.S.), China and India. We noted that temporal topic trends extracted from disease-related news reports successfully captured the dynamics of multiple outbreaks such as whooping cough in U.S. (2012), dengue outbreaks in India (2013) and China (2014). Our observations also suggest that efficient modeling of temporal topic trends using time-series regression techniques can estimate disease case counts with increased precision before official reports by health organizations.  
\end{abstract}
\begin{document}

\flushbottom
\maketitle
\thispagestyle{empty}

\section*{Introduction}

Infectious diseases are a threat to public health and economic stability of many countries. Open source indicators (e.g., news articles~\cite{brownstein2008surveillance,linge2009internet} , blogs~\cite{corley2010text} , search engine query volume~\cite{yuan2013monitoring,ginsberg2009detecting,santillana2014using,gu2015early} , social media chatter~\cite{denecke2012making,lee2013real,sugumaran2012real,paul2011you} and other sources~\cite{nsoesie2015monitoring}) are an attractive option for monitoring infectious disease progression, primarily due to their sheer volume and capacity to capture early signals of disease outbreaks, and in some cases, trends in population health-seeking behavior.  However, most prior work in digital surveillance using open source indicators has targeted specific diseases, such as influenza~\cite{nsoesie2015monitoring,chakraborty2014forecasting} and hantavirus pulmonary syndrome (HPS)~\cite{ghosh2015rare} . Therefore, there is a need to develop generic frameworks that are applicable to multiple infectious diseases.

Official surveillance reports released by health organizations (e.g., CDC, WHO, PAHO) are published with a considerable delay of weeks, months or even a year. Therefore, traditional surveillance systems are not always effective at real-time monitoring of emerging public health threats. Unlike traditional surveillance data, informal digital sources, such as news media, blogs, and micro-blogging sites (Twitter) are typically available in (near) real-time. Proper mining of signals from these digital sources can effectively help in minimizing the time lag between an outbreak start and formal recognition of an outbreak, allowing for an accelerated response to public health threats. The gains in supplementing traditional surveillance using digital sources have been discussed in Nsoesie et al.~\cite{nsoesie2015computational} , Salath{\'e} et al.~\cite{salathe2012digital,salathe2013influenza} and Hartley et al.~\cite{hartley2013overview}

Our key contributions are as follows. (i) We introduce {\fullmodel}, a generic temporal framework for analyzing disease-related news reports using a supervised topic model. The supervised topic model discovers multiple disease topics of interest and their associated temporal trends of prominence in news media. (ii) {\fullmodel} captures trends in disease progression, such as periodicity, peaks and troughs via temporal trends of disease topics in news media. (iii) {\fullmodel} also estimates disease incidence before official reports by health agencies using time-series regression models interposed over the temporal trends of disease topics.

We validated our method against disease case count reports, as available from public health agencies, in U.S., China and India.  Disease-related news articles were provided by HealthMap~\cite{freifeld2008healthmap} , an internationally recognized, global disease alert system capturing outbreak reports from over 200,000 electronic news sources. {\fullmodel} was evaluated on multiple outbreaks in the recent past, such as whooping cough in U.S. (2012)~\cite{cherry2012epidemic} , periodic outbreaks of avian influenza A(H7N9)~\cite{yang2014avian,gao2013clinical} and hand, foot, and mouth disease (HFMD) in China (2013 and 2014), periodic outbreaks of acute diarrheal disease (ADD) in India (2013 and 2014), major dengue outbreaks in China (2014)~\cite{shen2015multiple} and India (2013). Our experiments indicate that {\fullmodel} was successfully able to capture the dynamics of the mentioned outbreaks and estimate the case counts, before official reports were published. However, inconsistent news coverage was found to adversely affect the performance of our method in certain scenarios. 

\section*{Materials and Methods}

\subsection*{Data sources}
In this section, we discuss the data sources used to analyze the infectious disease outbreaks. We first describe the case count reports collected from public health agencies and complete our discussion about the HealthMap data used in this study.

\paragraph*{Disease case counts.} For each country, we collected case count data corresponding to multiple diseases over a certain time period.
In Table~\ref{tab:disease_case_count}, we show the disease names (along with methods of transmission), health agencies from which case counts were collected, time period over which case counts were obtained and temporal granularity (daily, monthly, weekly or yearly) of the obtained case counts corresponding to each country.

\begin{table*}[ht]
\centering
\begin{center}
\begin{tabular}{|l|l|l|l|l|}
\hline
Country & \makecell{Disease names \\ (Methods of transmission)} & \makecell{Health \\ agencies} & \makecell{Time \\ period} & \makecell{Temporal \\ granularity} \\
\hline
U.S. & \makecell{Whooping cough (airborne, direct contact) \\ Rabies (zoonotic) \\ Salmonellosis (food-borne) \\ E. coli infection (waterborne, food-borne)} & \makecell{Project Tycho~\cite{van2013contagious} \\ (\url{https://www.tycho.pitt.edu/})} & \makecell{January 2010 - \\ December 2013} & Weekly  \\
\hline
China & \makecell{H7N9 (zoonotic) \\ HFMD (direct contact, airborne) \\ Dengue (vector-borne)} & \makecell{National Health and \\ Family Planning Commission \\ (\url{http://en.nhfpc.gov.cn/})} & \makecell{January 2013 - \\ December 2014} & Monthly \\
\hline
India & \makecell{ADD (food-borne) \\ Dengue (vector-borne) \\ Malaria (vector-borne)} & \makecell{Integrated Disease \\ Surveillance Programme \\
(\url{http://www.idsp.nic.in/})} & \makecell{January 2013 - \\ December 2014} & Weekly \\
\hline
\end{tabular}
\end{center}
\caption{\label{tab:disease_case_count} \textbf{Disease names (along with routes of transmission), health agencies from which case counts were collected, time period over which case counts were obtained and temporal granularity (daily, monthly, weekly or yearly) of the obtained case counts corresponding to each country.} H7N9 stands for avian influenza A, ADD stands for acute diarrheal disease and HFMD stands for hand, foot, and mouth disease.}
\end{table*}

\paragraph*{HealthMap.} Disease-related news articles were found to be indicative of infectious disease outbreaks~\cite{ghosh2015rare} . We collected such articles related to the mentioned diseases in Table~\ref{tab:disease_case_count}, for each country under consideration, from HealthMap. The HealthMap corpus is a publicly available database from which we collected the disease-related articles, reported during the time period of interest. Each article contains the reported date and the corresponding location information in the form of (lat, long) co-ordinate pairs. We converted the location co-ordinates to location names (country, state) via reverse geocoding. Reverse geocoding is defined as the process of finding a readable address or place name for a given (lat, long) pair. For example, $(26.562851, -81.949532)$ was converted to (\textit{United States}, \textit{Florida}) after reverse geocoding. Each HealthMap article was passed through a series of preprocessing steps. For China, majority ($87.94 \%$) of the articles were published in either Traditional Chinese or Simplified Chinese. We translated the textual content of these articles to English for ease of analysis.  The articles were preprocessed by removing non-textual elements, tokenization~\cite{webster1992tokenization,singh2014effective} , lemmatization~\cite{kanis2010comparison} and removal of stop words via BASIS Technologies’ Rosette Language Processing (RLP) tools~\cite{naren2014forecasting,doyle2014forecasting} . For more details on these steps, see subsection `HealthMap preprocessing' within the section `Supplementary Information' at the end of the manuscript. The set of unique words in these processed articles were found to contain general- (e.g., \textit{cold}, \textit{contagious}, \textit{nausea}, \textit{blood}, \textit{food-borne}, \textit{waterborne}, \textit{sanitation}) as well as specific- (e.g., \textit{rabies}, \textit{whooping}, \textit{h7n9}, \textit{dengue}, \textit{salmonella}, \textit{malaria}) disease related terms. In Table~\ref{tab:article_word_numbers}, we show country-wise distribution of the total number of HealthMap news articles along with unique words and location names extracted from all the corresponding articles.

\begin{table*}[ht]
\centering
\begin{center}
\begin{tabular}{|l|l|l|l|}
\hline
Country & \makecell{Total number of HealthMap \\ news articles} & \makecell{Total number of unique words} & \makecell{Total number of unique location names \\ or (country, state) pairs} \\
\hline
China & 11,209 & 21,879 & 30 \\
\hline
India & 1,204 & 17,160  & 30 \\
\hline
U.S. & 9,872 & 59,687 & 51 \\
\hline
\end{tabular}
\end{center}
\caption{\label{tab:article_word_numbers} \textbf{Country-wise distribution of the total number of HealthMap news articles along with unique words and location names extracted from all the corresponding articles.}}
\end{table*}

Our next step was to extract the underlying topics related to the mentioned diseases in Table~\ref{tab:disease_case_count} and their associated temporal trends from the processed articles for each country. Following Rekatsinas et al.~\cite{ghosh2015rare} , the processed corpus for each country was transformed to a collection of tuples of the form $\lbrace w, l, t\rbrace: count$, where \textit{count} is the number of news articles mentioning the word $w$ associated with the location $l$ and time point $t$ in the tuple. For this transformation, we assumed that for each country, each processed article consists of words from a vocabulary $V$, corresponds to a discretized time window $t \in \lbrace1, 2, \cdots, T\rbrace$ and is geotagged with a location $l$ from a set of locations $L$ in the country. For China, disease case counts were available on a monthly granularity and as such each time point $t$ represents a period of 1 month. However, for diseases in U.S. and India, case counts were obtained on a weekly basis and as such time point $t$ represents a period of 1 week or more specifically, epidemiological week (hereafter referred to as epi week). For example, the tuple $(salmonella, (United States, Kansas), 2013-10-06): 9$ denotes that the word \textit{salmonella} was mentioned in 9 articles referring to the state of \textit{Kansas} in U.S. over the epi week extending from $6^{th}$ October 2013 to $12^{th}$ October 2013. For each country, let $N_{l}$ represent the collection of tuples for each location $l \in L$ and $\mathcal{X}$ denote the set of all tuple collections $N_{l}$ until time point $T$. This transformed set $\mathcal{X}$ was analyzed to extract the temporal trends of disease topics as discussed in the following section. Both $N_{l}$ and $\mathcal{X}$ were updated for each country, as we proceed along the time window.

\subsection*{{\fullmodel}}

In this section, we describe in details the components of our proposed framework {\fullmodel}. The first component is the supervised topic model used to extract temporal topic trends from $\mathcal{X}$. The second component, referred to as {\fullmodelforecast}, is responsible for generating estimates of disease case counts using past available case counts and temporal topic trends extracted by the supervised topic model. 

\subsubsection*{Temporal topic modeling}

The first component of {\fullmodel} deals with the topic and pattern discovery problem. The set $\mathcal{X}$ of all tuple collections $N_{l}$ can be treated as a three-dimensional matrix of size $V \times L \times T$ where the dimensions are represented by words (size $V$), locations (size $L$) and time points (size $T$). Each element $x_{w,l,t}$ in $\mathcal{X}$ represents the total number of articles mentioning the word $w$ ($w \in V$) referring to location $l$ ($l \in L$) over the time point $t$ ($t \in {1,2,3,...,T}$). We assume that each entry in a non-zero element $x_{w,l,t}$ of $\mathcal{X}$ is associated with a latent disease topic and therefore, such hidden disease topics can be modeled in terms of three dimensions of $\mathcal{X}$. Our goal is to extract the hidden disease topics and their corresponding associations with each dimension of $\mathcal{X}$. Following previous literature on topic models~\cite{blei2003latent,blei2008supervised,ghosh2015rare,jagarlamudi2012incorporating} , we implemented a supervised temporal topic model for this purpose. We supervise the discovery process of each disease topic by providing a set of prior words (also called seed words)~\cite{jagarlamudi2012incorporating} . These seed words are user-provided prior knowledge of each infectious disease and they encourage the topic model to find evidence of these disease topics in the HealthMap corpus. This supervised method helps in improving the discovery of word co-occurrences within each topic as the model tends to discover words that are related to the words in the seed set. Additionally, we model time and location jointly~\cite{ghosh2015rare} with the word co-occurrence patterns. This enables tracking of temporal and spatial patterns of these disease topics in the news.  For more details on the supervised topic model, see subsection `Generative process of the supervised topic model' within the section `Supplementary Information' at the end of the manuscript.

The supervised topic model takes $\mathcal{X}$ as input, discovers $K$ disease topics and decomposes $\mathcal{X}$ into four two-dimensional matrices as shown below. Each two-dimensional matrix represents the association between the discovered disease topics and the dimensions in $\mathcal{X}$. 

\begin{itemize}
\item $\xi$: A $K \times T$ matrix where each row represents a discrete probability distribution over the time points ($1, 2, 3, ....., T$) for a specific topic $z \in {1, 2, 3, ...., K}$. Each row of $\xi$ ($\xi_{z}$) represents the temporal topic trends or distribution for the disease topic $z \in {1, 2, 3, ...., K}$.

\item $\phi^s$: A $K \times S$ matrix where each row represents a discrete probability distribution over the set $S$ of seed words for a specific topic $z \in {1, 2, 3, ...., K}$. $\phi^s$ is hereafter referred to as the seed topic distribution.

\item $\phi^r$: A $K \times V$ matrix where each row represents a discrete probability distribution over the set of regular words for a specific topic $z \in {1, 2, 3, ...., K}$. The set of regular words refers to all the words in vocabulary $V$ including the seed words. $\phi^r$ is hereafter referred to as the regular topic distribution.

\item $\theta$: A $L \times K$ matrix where each row represents a discrete probability distribution over $K$ topics for a specific location $l \in L$. 

\end{itemize}

For more details on $\xi$, $\phi^s$, $\phi^r$ and $\theta$, see subsection `Generative process of the supervised topic model' within the section `Supplementary Information' at the end of the manuscript.

\paragraph*{Inference.} To compute the output parameters $\theta$, $\phi^r$ , $\phi^s$ and $\xi$ in the supervised topic model given input observed data $\mathcal{X}$, we need to solve an inference problem. In topic models, exact computation is intractable~\cite{blei2003latent} and thus we are interested in approximate inference of the model parameters. Since collapsed gibbs sampling~\cite{steyvers2007probabilistic,matsubara2012fast,porteous2008fast} is a straight-forward, easy to implement, and unbiased approach that converges rapidly to a known ground-truth, it is typically preferred over other possible approaches~\cite{blei2003latent,minka2002expectation} in large scale applications of topic models~\cite{ghosh2015rare,matsubara2012fast,rosen2004author} . Thus we used collapsed gibbs sampling as the inference scheme for the supervised topic model. For more details on the inference process, see subsection `Inference via collapsed gibbs sampling' within the section `Supplementary Information' at the end of the manuscript.

\paragraph*{Seed word extraction.} Seed words for each disease topic were extracted by examining the content of a subset of news articles mentioning the disease. Additionally, following similar techniques as in Chakraborty et al.~\cite{chakraborty2014forecasting} , we also examine a number of expert websites, such as CDC and WHO, to identify the most important keywords for a particular disease. Seed words used in this study are shown in Tables 5, 6 and 7 corresponding to diseases in U.S., China and India respectively.

\subsubsection*{Estimation of disease case counts} 

The second component of {\fullmodel} is concerned with estimation of disease case counts using relevant information such as past case counts and temporal topic trends ($\xi$). Let $D$ be the disease of interest. Without loss of generality, let the $z^{th}$ disease topic corresponds to $D$. Furthermore, let $S_{D,T}$ denotes case counts of $D$ and $\xi_{z,T}$  denotes temporal trend value for $z^{th}$ disease topic at a time point $T$. In general, reports of case counts published by health organizations are delayed (see Chakraborty et al.~\cite{chakraborty2014forecasting}, Wang et al.~\cite{wang2015dynamic}) and hence, at time point $T$ case counts are available only till $T' < T$ with a delay $\delta=T-T'$. However, temporal topic trend values ($\xi_{z,1}, \xi_{z,2}, \cdots, \xi_{z,T}$) are available till T. Hence, we can formally define the case count estimation problem as estimating $S_{D,T}$ using past case counts ($S_{D}$) available till $T'$ and temporal topic trends ($\xi_{z}$) available till $T$. In general, disease case counts have a publication delay of 1 time point ($T' = T - 1$) and hence, estimating $S_{D,T}$ at $T$ is equivalent to 1-step ahead estimation. 

\paragraph*{\fullmodelforecast.} For 1-step ahead case count estimation, we used a regularized version of autoregressive model with external input variables (ARX) where external input variables are represented by the temporal topic trends ($\xi_{z}$). We used Elastic Net~\cite{zou2005regularization} as the regularization model in ARX. This estimating component of {\fullmodel} is designated as {\fullmodelforecast} and defined below in equation (\ref{eq:ari}).

\begin{equation}
  \label{eq:ari}
  \hat{S}_{D,T} = \underset{\mbox{Internal component}}{\underbrace{\sum_{i=1}^p{\gamma_iS_{D,T-i}}}}
  + \underset{\mbox{External component}}{\underbrace{\sum_{j=1}^{q}{\eta_j g_r\left(\xi_{z,T-j + 1 +s}\right)}}}
  +  \epsilon_{D, T} 
\end{equation}

where, $\hat{S}_{D,T}$ is the estimated case count for disease $D$ at time point $T$ and $\gamma_i, \eta_j$ are the regression coefficients fitted using Elastic Net constraints as given below in equation (\ref{eq:ari:fit}).

\begin{equation}
  \label{eq:ari:fit}
  \begin{array}{lcl}
  \gamma_{opt},\ \eta_{opt}  & = & \mbox{arg}\min\limits_{\gamma, \eta} 
  \ \sum\limits_{t'=0}^{T'} \left({S_{D,t'}} - \hat{S}_{D,t'}\right)^2 
  + \lambda_1\sum_{i,j}|\gamma_i + \eta_j |
  + \lambda_2\sum_{i,j}\left(\gamma_i + \eta_j \right)^2
  \end{array}
\end{equation}

where, $\lambda_1$ and $\lambda_2$ are the regularization coefficients for the $L1$ and $L2$ components of Elastic Net, respectively. The Elastic Net combines the properties of Least Absolute Shrinkage and Selection Operator (LASSO)~\cite{tibshirani1996regression,hastie2009elements} and Ridge regression~\cite{hastie2009elements} models. This combination allows for learning a sparse model like LASSO, while still maintaining the regularization properties of Ridge. If $\lambda_1$ equals to 0, equation (\ref{eq:ari:fit}) equates to a Ridge estimator. On the other hand, if $\lambda_2$ equals to 0, equation (\ref{eq:ari:fit}) corresponds to a LASSO estimator.

There are broadly two components to equation (\ref{eq:ari}) which captures different signals about the diseases as follows. (i) \textbf{Internal component ($p$):} This component is an autoregressive model that captures the signal embedded in past case counts and thus describes a delayed model. $p$ indicates the order of autoregression. (ii) \textbf{External component ($q$, $r$, $s$):} This component can also be thought of as an autoregressive component over the temporal topic trends ($\xi_{z}$) where $q$ is the number of time points to look back. The temporal topic trends are subjected to two additional transformations as follows. (a) \textbf{Shift indicator ($s$):} Often, the incidence of news reports is not concurrent with the incidence of diseases, as recorded in the case counts. {\fullmodelforecast} incorporates this information by shifting the temporal topic trend value $\xi_{z,T}$ by $s$ steps. The shift can be positive (indicating a lagging trend), negative (indicating a leading trend) or zero (indicating a co-incident trend). (b) \textbf{Rolling transformation ($r$):} Disease case counts ($S_{D}$) do not follow a strictly linear relationship with temporal topic trends ($\xi_{z}$). One of the simplest methods is to detrend the signals using difference of trend values instead of absolute values. However, our experiments showed that such transformations using a single time point often lead to unstable estimates. As such, we define a rolling transformation $g$ over a window length $r$ given below in equation (\ref{eq:rolling_transform}). 

\begin{equation}
\label{eq:rolling_transform}
	g_r(x_T) = x(T) -  x(T -r)
\end{equation}

Essentially, such transformations aim to capture the changes in trend values over a period and were found to be more indicative than absolute values. We ran a cross-validation step to find the optimal ($p$, $q$, $r$, $s$) parameters.

\paragraph*{Converting temporal topic trends to sampled case counts.} We described {\fullmodelforecast} using the temporal topic trends or distribution ($\xi_{z}$) as the external input variables. It is to be noted that the disease case counts ($S_{D}$) and the temporal topic distribution ($\xi_{z}$) are typically at  different numerical scales since values in a distribution range from 0 to 1. To improve numerical stability we converted the temporal topic distributions to estimated case counts using multinomial sampling~\cite{kerns2010introduction} over the time range. In multinomial sampling, samples are drawn from a multinomial distribution~\cite{kerns2010introduction} . The case counts estimated via multinomial sampling from the temporal topic distributions are hereafter referred to as sampled case counts. To calculate the sampled case counts ($\Xi_{D}$) for disease $D$, the corresponding temporal distribution $\xi_{z}$ for $z^{th}$ topic was used as the multinomial distribution and the total number of case counts available till $T' < T$ at $T$ (due to delay in reporting of case counts) was used as the number of samples to be drawn from the distribution. See Algorithm (\ref{al:sampling}) for more details.

\begin{algorithm}
  \DontPrintSemicolon
  \SetKwInOut{Input}{Input}\SetKwInOut{Output}{Output}  
  \Input{Temporal topic distribution: $\xi_{z,1}, \dots, \xi_{z, T}$\\
  Total number of case counts till time point $T'$: $TS_{D,T'} = \sum\limits_{t'=0}^{T'}(S_{D,t'})$
  }
  \Output{Sampled case counts from temporal topic distribution: $\Xi_{D,1}, \dots, \Xi_{D, T}$}
  $p \leftarrow \xi_{z,1}, \dots, \xi_{z, T}$\\
  $n \leftarrow TS_{D,T'}$\\
  Draw $n$ time points $0 \leq t_s \leq T$ using multinomial sampling where $p$ is the multinomial distribution and $n$ is the total number of samples to be drawn. \\
  For each time point $0 \leq t_s \leq T$, sampled case count $\Xi_{D,t_s}$ is calculated as the frequency of occurrence of $t_s$ in the above $n$ number of samples (time points) drawn from the multinomial distribution $p$.
  \caption{Multinomial sampling to convert temporal topic distribution to sampled case counts. \label{al:sampling}}
\end{algorithm}

\section*{Results}

In this section, we present an empirical evaluation of our proposed framework {\fullmodel}.
We first evaluated the disease topics discovered by the supervised topic model. Next, we analyzed whether the temporal topic trends ($\xi$) extracted by the supervised topic model are able to capture disease dynamics - including seasonality, abrupt peaks and troughs. Finally, we evaluated the quality of case counts estimated by {\fullmodelforecast} against the actual disease case counts.

\subsection*{Disease topic discovery}

To evaluate the discovered disease topics, we looked at the words having higher probabilities in the seed topic distributions ($\phi^s$) and regular topic distributions ($\phi^r$). We present the analysis of $\phi^s$ and $\phi^r$ in Tables 5, 6 and 7 corresponding to disease topics in U.S., China and India respectively. For each country, both $\phi^s$ and $\phi^r$ were extracted from HealthMap data spanning over the entire time period shown in Table~\ref{tab:disease_case_count}. For each disease topic ($z$), we show the seed words and their corresponding probabilities (sorted in descending order) in the seed topic distribution $\phi^s_{z}$. Seed words having higher probabilities in $\phi^s_{z}$ serve as informative prior words in the topic discovery process as they are mentioned frequently in news articles related to the $z^{th}$ disease topic. For example, seed words such as \textit{food}, \textit{salmonella}, \textit{product}, \textit{fda}, \textit{drug}, \textit{contamination} serve as informative prior words for the discovery of salmonellosis topic in U.S. since they have higher probabilities in the seed topic distribution (see Table 5). On the other hand, seed words such as \textit{enteritidis}, \textit{newport} provide less prior information due to their low probability values in the seed topic distribution. To understand how the supervised topic model discovers words from the HealthMap corpus related to these input seed words, we also show some of the regular words having higher probabilities in the regular topic distribution $\phi^r_{z}$. For a particular disease topic, these regular words with higher probabilities are mentioned frequently in news articles related to that disease and also capture different aspects (causes and clinical symptoms, methods of transmission, etc.) of the disease that the topic represents. For example, in Table 5 we show these regular words (having higher probabilities in the regular topic distribution $\phi^r_{z}$) for the salmonellosis topic in U.S. Words such as \textit{diarrhea}, {nausea}, \textit{vomit} are related to clinical symptoms of salmonellosis. On the other hand, words such as \textit{eat}, \textit{contaminated}, \textit{restaurant}, \textit{meat}, \textit{beef} are related to causes of salmonellosis. 

\subsection*{Detection of outbreak patterns}

We also examined the temporal distribution or trends ($\xi_{z}$) for each disease topic ($z$) in a specific country (Figures 1, 2 and 3) and their correlations with the disease case counts. For each country, temporal topic trends ($\xi_{z}$) were extracted from HealthMap data spanning over the entire time period shown in Table~\ref{tab:disease_case_count}. We made several important observations as follows. 

\paragraph*{\bf Disease seasonality.} In U.S., case counts of salmonellosis and E. coli infection exhibit strong periodic outbreaks, both peaking during the summer (see Figures 1 (e) and (g)). Temporal topic trends extracted by {\fullmodel} were able to capture the periodicity of these two diseases, particularly periodic outbreaks of salmonellosis and E. coli infection in 2010, 2012 and 2013. However, during 2011, temporal topic trends failed to monitor the peak season properly though they show a tendency to increase during summer. For salmonellosis in 2013, the temporal topic trends captured the major peak of the outbreak at the start of the season while failing to capture the seasonal activity towards the end. For rabies, although the topic trends captured the general characteristics it failed to detect some major outbreaks, such as the outbreak in the summer of 2010 (see Figure 1 (c)). 

In China, H7N9 and HFMD case counts exhibit strong periodic outbreaks, with H7N9 peaking during the winter and HFMD peaking during the summer (see Figures 2 (a) and (c)). For H7N9, temporal topic trends extracted by {\fullmodel} were able to detect the seasonal outbreaks during March-April 2013 and January-February 2014. However, for HFMD, peaks in temporal topic trends precede the peaks in case counts during the summer of 2013 and 2014 respectively. Therefore, temporal topic trends for HFMD exhibit a negative shift (leading indicator) with respect to the case counts.

In India, case counts of ADD exhibit periodic outbreaks, peaking during the summer of 2013 and 2014 (see Figure 3 (a)). Temporal topic trends detected the seasonal outbreak in the summer of 2013 but failed to capture the outbreak in the summer of 2014. 

\paragraph*{\bf Sudden peaks/troughs.} In U.S., whooping cough outbreaks do not exhibit yearly periodicity unlike salmonellosis and E. coli infection (see Figure 1 (a)). There was a major outbreak of whooping cough during the summer of 2012 and {\fullmodel} detected this sudden increase (peak) in case counts by displaying higher topic trends during the entire period of the outbreak. {\fullmodel} also did not detect outbreaks during periods (summer of 2011 and 2013) known to have low incidences (troughs) of whooping cough by displaying lower topic trends, suggesting low false alarm rate. 

In China and India, dengue case counts exhibit seasonal outbreaks with peaks in case counts appearing during the months of September and October. However, China experienced a severe dengue outbreak in 2014~\cite{shen2015multiple} in comparison to the outbreak in 2013 with the peak value of case counts exceeding 25,000 in the month of October (see Figure 2 (e)). Temporal topic trends detected this sudden massive increase in case counts by displaying a sharp spike during the outbreak period. India also experienced a large dengue outbreak in 2013 with the peak value of case counts exceeding 3,000 during a particular epi week in October (see Figure 3 (c)). {\fullmodel} was able to detect this outbreak by displaying higher topic trends during the peak period. Malaria case counts in India exhibit irregular outbreaks or peaks (see Figure 3 (e)). {\fullmodel} was successful in capturing majority of these outbreaks though it failed to detect some major peaks, such as the peak during the month of June 2014.

\paragraph*{\bf Sampled case counts.} Along with the temporal topic trends ($\xi_{z}$), we also showed the corresponding sampled case counts ($\Xi_{D}$) generated via multinomial sampling (see Algorithm (\ref{al:sampling})) from $\xi_{z}$ for a disease $D$ in Figures 1 ((b), (d), (f) and (h)), 2 ((b), (d) and (f)), 3 ((b), (d) and (f)).  The figures show that the sampled case count values share similar numerical range as the disease case counts while maintaining shapes of the temporal topic trends. On the other hand, the temporal topic trend values are at different numerical range (ranging from 0 to 1) with respect to the case counts.

\subsection*{Estimating case counts}

As official reports of case counts by health agencies are usually lagged by a single time point (week or month), reliable early estimates of disease incidence can facilitate the allocation of public health resources to enable effective control measures. Therefore, we aim to perform 1-step ahead estimation of disease case counts starting from a particular time point (For definition of 1-step ahead estimation, see subsection `Estimation of disease case counts' within section `{\fullmodel}' of `Materials and Methods'). For the purpose of experimental validation, we used historical HealthMap data over a certain time period as the static training set in a specific country (referred to as the static training period) and progressively utilized the remaining time points as the evaluation period over which we evaluated the case count estimates of {\fullmodelforecast}. To estimate case counts at a particular time point $T$ within the evaluation period, we utilized HealthMap data from $t=0$ up to $t=T$ and extracted disease topics using the supervised topic model. The disease case counts at $T$ were next estimated using past case counts available up to $t=T'$ ($T' = T-1$) and temporal topic trends (or, sampled case counts) available up to $t=T$. In Table~\ref{tab:time_periods}, we show the total time period of study, static training period and the evaluation period for each country. 

\paragraph*{Baselines.} For the task of 1-step ahead estimation, we compared the performance  of {\fullmodelforecast} against 2 baseline methods, namely {\fullmodelbaseline} and {\baselinemodel}. In {\baselinemodel}, we fitted an autoregressive-moving-average model (ARMA~\cite{box2011time}) over past disease case counts to generate case count estimates. {\baselinemodel} doesn't use any information related to temporal topic trends ($\xi_{z}$). However, in case of {\fullmodelbaseline}, we used an autoregressive–moving-average model with external input variables (ARMAX~\cite{box2011time}) where external input variables incorporate the information embedded in temporal topic trends. For more details on the baseline methods, see subsection `Baseline methods for case count estimation' within the section `Supplementary Information' at the end of the manuscript. We also compared temporal topic trends against sampled case counts (generated by multinomial sampling from the temporal topic trends) as the external input variables, for the applicable methods {\fullmodelforecast} and {\fullmodelbaseline}. 

\paragraph*{Evaluation.} We evaluated the case count estimates of each method over the evaluation period by comparing them against the actual case counts using normalized root-mean-square error (NRMSE). In Table~\ref{tab:case_count_estimate}, we present a comparative performance evaluation of the methods for 1-step ahead estimation in terms of NRMSE values corresponding to diseases in U.S., China and India respectively. Table~\ref{tab:case_count_estimate} provides multiple insights as follows. (i) {\fullmodelforecast} with sampled case counts as external variables is the best performing method achieving lowest NRMSE values for majority (8 out of 10) of the $\lbrace$country, disease$\rbrace$ combinations. (ii) Two exceptions are $\lbrace$China, HFMD$\rbrace$ and $\lbrace$U.S., E. coli infection$\rbrace$ where {\fullmodelforecast} and {\fullmodelbaseline} with temporal topic trends as external variables achieve lowest NRMSE values respectively. (iii) Both {\fullmodelforecast} and {\fullmodelbaseline} perform better overall with sampled case counts as external variables than temporal topic trends. (iv) For none of the $\lbrace$country, disease$\rbrace$ combinations, {\baselinemodel} is able to achieve lowest NRMSE values indicating the significance of incorporating temporal topic trends or sampled case counts as external variables for estimating case counts.        

\begin{table*}[ht]
\centering
\begin{center}
\begin{tabular}{|l|l|l|l|}
\hline
Country & \makecell{Total time period \\ of study} & \makecell{Static training \\ period} & \makecell{Evaluation \\ period} \\
\hline
U.S. & \makecell{January 2010 - December 2013} & \makecell{January 2010 - December 2011} & \makecell{January 2012 - December 2013} \\
\hline
China & \makecell{January 2013 - December 2014} & \makecell{January 2013 - March 2013} & \makecell{April 2013 - December 2014} \\
\hline
India & \makecell{January 2013 - December 2014} & \makecell{January 2013 - November 2013} & \makecell{December 2013 - December 2014} \\
\hline
\end{tabular}
\end{center}
\caption{\label{tab:time_periods} \textbf{Total time period of study, static training period and the evaluation period for estimating disease case counts in each country.}}
\end{table*}

\begin{table*}[ht!]                                                               
  \centering                                                                     
  \resizebox{\textwidth}{!}{%
  \begin{tabular}{|c|c|c|c|c|c|c|}                                                
  \hline                                                                        
  \multirow{2}{*}{Country} & \multirow{2}{*}{Disease} & \multirow{2}{*}{{\baselinemodel}} & \multicolumn{2}{c|}{{\fullmodelbaseline}} & \multicolumn{2}{c|}{{\fullmodelforecast}} \\ \cline{4-7}
                        &                          &                                & \makecell{with temporal \\ topic trends}      & \makecell{with sampled \\ case counts}      & \makecell{with temporal \\ topic trends}  & \makecell{with sampled \\ case counts} \\ \hline
  \multirow{4}{*}{U.S.}    & {\it \makecell{Whooping \\ cough}}   & 0.584                          & 0.577              & 0.582           & 0.583          & {\bf 0.558}  \\ \cline{2-7}
                           & {\it Rabies}          & 0.875                          & 0.888              & 0.886           & 0.877          & {\bf 0.865}  \\ \cline{2-7}
                          & {\it Salmonellosis}    & 0.445                          & 0.978             & 0.450           & 0.441          & {\bf 0.430}  \\ \cline{2-7}
                          & {\it \makecell{E. coli \\ infection}}  & 0.685                          & {\bf 0.657}        & 0.663           & 0.686          & 0.671  \\ \hline
  \multirow{4}{*}{China}     & {\it H7N9}             & 1.096                          & 0.850              & 0.888           & 1.027          & {\bf 0.712}  \\ \cline{2-7} 
                        & {\it HFMD}      & 1.574                          & 1.524             & 1.538           & {\bf 0.622}          & {0.626}  \\ \cline{2-7}
                        & {\it Dengue}  & 1.076                          & 0.639        & 0.634           & 1.094          & {\bf 0.549}  \\ \hline
 \multirow{4}{*}{India}  & {\it ADD}     & 1.226                          & 1.285              & 1.119           & 0.844          & {\bf 0.833}  \\ \cline{2-7} 
                        & {\it Dengue}  & 0.966                          & 1.086              & 1.021           & 1.073          & {\bf 0.878}  \\ \cline{2-7} 
                        & {\it Malaria} & 1.060                          & 1.062              & 1.047           & 1.016          & {\bf 0.963}  \\ \hline
 \end{tabular}                                                                 
 }                                                                 
 \caption{\label{tab:case_count_estimate} \textbf{Comparing the performance of {\fullmodelforecast} against the baseline methods {\fullmodelbaseline} and {\baselinemodel} for 1-step ahead estimation of disease case counts.} Metric used for comparing the case counts estimated by the methods against the actual case counts is the normalized root-mean-square error (NRMSE).}
 \end{table*}

\section*{Discussion}

In this paper, we studied the problem of monitoring and estimating outbreaks of multiple infectious diseases using disease-related online news reports obtained from HealthMap. We introduced {\fullmodel}, a novel and generic temporal framework that combines supervised temporal topic models with time-series regression techniques to monitor and estimate disease incidence. Experimental results demonstrate that {\fullmodel} is able to capture the time varying incidence of multiple diseases via temporal topic trends. Our experiments also illustrate that {\fullmodel} can estimate disease incidence 1-step ahead with increased accuracy using information from temporal topic trends.

{\fullmodel} uses online news reports as the sole data source to capture disease dynamics during outbreaks. Therefore, it is generic in the sense that it is not tailored to a particular disease or class of diseases. Moreover, the set of diseases selected for each country represent a diversity of transmission pathways as shown in Table~\ref{tab:disease_case_count}. Hence, the applicability of {\fullmodel} to these diverse sets of diseases as demonstrated in this study showcases the potential generalizability of our approach to different class of diseases. 

Temporal topic trends extracted by {\fullmodel} from HealthMap news reports successfully captured dynamics of multiple outbreaks, such as whooping cough in U.S. during summer of 2012, periodic outbreaks of salmonellosis and E. coli infection in U.S., periodic outbreaks of H7N9 and HFMD in China, dengue outbreaks in India (2013) and China (2014). However, there are certain deviations where temporal topic trends could not monitor the trends in disease outbreaks properly, such as salmonellosis and E. coli infection outbreaks in 2011, rabies outbreak in 2010 and ADD outbreak in 2014. We posit that such deviations are a factor of news media coverage during disease outbreaks, which is driven by interest. Moreover, our framework is heavily reliant on news-corpora and does not account for possible reporting errors. As such inconsistent interest-driven news coverage and articles with missing content affect the performance of our framework. 

{\fullmodel} supports monitoring and also 1-step ahead estimation of disease case counts with increased precision. Table~\ref{tab:case_count_estimate} shows that {\fullmodelforecast} yields lowest NRMSE values for majority of the diseases when compared to the baseline methods {\fullmodelbaseline} and {\baselinemodel}. This implies that incorporating information from temporal topic trends via {\fullmodelforecast} results in improved estimation of case counts. It is also to be noted that {\fullmodelforecast} with sampled case counts as external variables achieves lower NRMSE for most of the diseases than the variant using temporal topic trends. This validates our claim that using sampled case counts instead of actual topic trends as the external variables adds numerical stability to {\fullmodelforecast}. However, {\fullmodelbaseline} is not able to provide significant performance improvement over {\baselinemodel}, which highlights the limitations of using ARMAX models in our framework for estimating case counts.

For dengue and HFMD in China, {\fullmodelforecast} shows considerable improvement on 1-step ahead estimation of disease incidence when compared to the baselines, specifically {\baselinemodel} (see Table~\ref{tab:case_count_estimate}). In order to have a clearer understanding of the improved performance of {\fullmodelforecast} with respect to the baselines, we plotted the temporal correlation between actual case counts and case counts estimated by the methods in Figure 4 corresponding to dengue and HFMD in China. It can be observed that {\fullmodelforecast} with sampled case counts as external variables is able to estimate the peak in dengue case counts more accurately in comparison to the baselines (see Figure 4 (a)). For HFMD, {\fullmodelforecast} with both topic trends and sampled case counts as external variables are able to estimate the peak in case counts, while the baselines fail to do so (see Figure 4 (b)). {\baselinemodel}'s inability to estimate the peaks in case counts for both dengue and HFMD implies that past case counts are not reliable indicators for estimating sudden increases or peaks in disease incidence and therefore, need to be augmented with disease signals from online news media for accurate estimation of outbreaks. However, inconsistent news coverage can adversely affect the timely estimation of outbreaks by {\fullmodelforecast} as shown in Figure 4 (c). India experienced periodic outbreaks of ADD with peaks in case counts during the summer of 2013 and 2014. However, we observe a lack of news coverage (no peak in temporal topic trends) during the peak in 2014 compared to the peak in 2013 (see Figures 3 (a) and (b)). Therefore, the case count estimates generated by {\fullmodelforecast} have a delayed peak with respect to the actual peak in case counts during the outbreak in 2014 (see Figure 4 (c)). This delayed peak is due to the internal component ($p$) in equation (\ref{eq:ari}) which extracts information from past case counts. 

Additional studies can focus on adapting {\fullmodel} to news-corpora inconsistencies by leveraging information from other sources, such as climatic attributes (temperature~\cite{akil2014effects} , precipitation~\cite{curriero2001association} , humidity~\cite{hales2002potential}) for calibration purposes. However, observations in this study suggest that monitoring progression of infectious diseases is possible and disease incidence can be estimated with increased precision via efficient capturing of signals from online news media.

\bibliography{sample}

\section*{Acknowledgements}

Saurav Ghosh, Prithwish Chakraborty, Sumiko R. Mekaru, John S. Brownstein and Naren Ramakrishnan, are supported by a research grant from the Intelligence Advanced Research Projects Activity (IARPA) via Department of Interior National Business Center (DoI/NBC) contract number D12PC000337. Elaine O. Nsoesie is supported by funding from the National Institute of Environmental Health Sciences of the National Institutes of Health (Award Number K01ES025438). The US Government is authorized to reproduce and distribute reprints for Governmental purposes notwithstanding any copyright annotation thereon. The views and conclusions contained herein are those of the authors and should not be interpreted as necessarily representing the official policies or endorsements, either expressed or implied, of IARPA, DoI/NBC, or the US Government.



\section*{Supplementary Data}


The data set used in this paper can be found in \url{https://github.com/sauravcsvt/EpiNews_supplementary_data}.

\section*{Supplementary Information}

\subsection*{HealthMap preprocessing}

Each HealthMap article was preprocessed using the following techniques.

\begin{enumerate}

\item \textbf{Removing non-textual elements.} We extracted the main textual content of each article using Dragnet~\cite{peters2013content} and Goose (\url{https://github.com/grangier/python-goose}), ignoring the non-textual elements such as images within the article.

\item \textbf{Tokenization and lemmatization.} Tokenization~\cite{webster1992tokenization,singh2014effective} is the process of segmenting a textual content into words, phrases, symbols or other meaningful elements commonly referred to as tokens. Lemmatization~\cite{kanis2010comparison} is performed after tokenization and can be defined as the normalization process in which various inflected forms of a word are converted to the same underlying lemma so that they can be analyzed as a single term. For example, terms such as \textit{travel}, \textit{traveled}, \textit{travels}, \textit{TRAVEL}, \textit{traveling}, \textit{Travelling}, \textit{travelling}, \textit{travelled}, \textit{Travel}, \textit{Traveling} were converted to the same underlying lemma \textit{travel}. Both tokenization and lemmatization were performed on the extracted textual content using  BASIS Technologies’ Rosette Language Processing (RLP) tools~\cite{naren2014forecasting,doyle2014forecasting} to generate a set of unique words or phrases corresponding to each article. 

\item \textbf{Uppercase to lowercase.} In this step, we converted the uppercase letters in each extracted word to lowercase letters. For example, both the terms \textit{Salmonella} and \textit{salmonella} convey the same meaning, so they were converted to a single term \textit{salmonella}. 

\item \textbf{Removal of stop words.} In the final step, we removed all the stop words such as \textit{in}, \textit{by}, \textit{of}, \textit{at}, \textit{all}, etc. from the set of unique words or phrases extracted from each article.

\end{enumerate}

\subsection*{Generative process of the supervised topic model}

In this section, we will discuss in details the generative process of the supervised topic model. Before going into the details of the generative process, we will first define the notion of a topic in the supervised topic model. In unsupervised topic models~\cite{blei2003latent,ghosh2015rare,matsubara2012fast,rosen2004author} , each topic $k$ is defined as a discrete probability distribution over all the words in the vocabulary $V$. In the supervised topic model, the notion of a topic is extended and defined as the convex combination of two discrete probability distributions: seed topic distribution and regular topic distribution~\cite{jagarlamudi2012incorporating} . The seed topic distribution can only generate words from the seed set $S$, and thus it is defined as a discrete probability distribution over only the words in the seed set $S$. On the other hand, the regular topic distribution has the freedom to generate any word including the seed words. So a regular topic is defined as a discrete probability distribution over all the words in the vocabulary $V$. Here we assume that each regular topic is associated with only one seed topic, i.e., there is a one-to-one correspondence between seed and regular topics. 

\begin{algorithm}
  \normalsize
  \DontPrintSemicolon
  \For{each topic $k \leftarrow \lbrace 1, 2, \dots, K \rbrace$}{
        Draw $\phi^{r}_{k}\ \sim\ \mbox{Dirichlet}(\beta^{(k)})$\;
        Draw $\phi^{s}_{k}\ \sim\ \mbox{Dirichlet}(\mu^{(k)})$\;
        Draw $\xi_{k}\ \sim\ \mbox{Dirichlet}(\gamma^{(k)})$\;
        Draw $\pi_{k}\ \sim\ \mbox{Beta}(1,1)$\;
  }
  \For{each location $l \in L$}{
    Draw $\theta_l\ \sim\ \mbox{Dirichlet}(\alpha^{(l)})$\;
    \For{each entry $i \in N_l$}{
      Draw topic $z_i \sim \mbox{Discrete}\left(\theta_l\right)$\;
      Draw indicator variable $x_i\sim\mbox{Bernoulli}\left(\pi_{z_i}\right)$\;
      Draw $w_i \sim \begin{cases}\begin{array}{lll}\mbox{Discrete}\left(\phi^{r}_{z_i}\right)  & \mbox{\sf when }x_i = 0 & \mbox{  {\sf// regular topic}}\\
        \mbox{Discrete}\left(\phi^{s}_{z_i}\right)  & \mbox{\sf when }x_i = 1& \mbox{ {\sf// seed topic}}\\
      \end{array}\end{cases}$\;
      Draw timestamp $t_i\sim \mbox{Discrete}\left(\xi_{z_i}\right)$\;
    }
  }
  \caption{Generative process of the supervised topic model\label{gen_topic}}
\end{algorithm}

The generative process of the supervised topic model is described in Algorithm (\ref{gen_topic}). Given $K$ disease topics, $L$ locations and $N_{l}$ for each $l \in L$, the supervised topic model uses location and topic specific discrete probability distributions to model the generation of word and time point in each entry of $N_{l}$. To generate each entry $i \in N_{l}$ for a location $l \in L$, we first sample a topic $z_{i}$ ($z_{i} \in \lbrace 1,2,\cdots,K\rbrace$) from the location-specific discrete probability distribution $\theta_{l}$ over $K$ disease topics. To generate a word $w_{i}$, we choose either the seed topic distribution ($\phi^{s}$) or the regular topic distribution($\phi^{r}$) corresponding to the sampled topic $z_{i}$. The indicator variable $x_{i}$ sampled from Bernoulli ($\pi_{z_{i}}$) decides whether the word should be drawn from the seed topic distribution or the regular topic distribution. $\pi_{z_{i}}$ is called the sampling probability for topic $z_{i}$. Once the distribution is chosen, the word $w_{i}$ is generated from it. Finally, the time point $t_{i}$ is drawn from the topic-specific discrete probability distribution $\xi_{z_{i}}$ over the time points $\lbrace1,2,\cdots,T\rbrace$. 

\paragraph*{Choice of priors.} $\xi_{k}$ ($k \in \lbrace1,2,....,K\rbrace$) is drawn from an asymmetric Dirichlet prior~\cite{rubin2012statistical,wallach2009rethinking} parameterized by a $T$-dimensional vector $\gamma^{(k)}$ as defined below in equation (\ref{eq:dirichlet_gamma}).

\begin{equation}
  \label{eq:dirichlet_gamma}
  \gamma^{(k)} = [N_{k,1} + \gamma^{'}, N_{k,2} + \gamma^{'},\cdots, N_{k,t} + \gamma^{'},\cdots, N_{k,T} + \gamma^{'}]
\end{equation}

where, $N_{k,t}$ is the sum of the \textit{count} variable across those tuples ($\lbrace w, l, t^{'}\rbrace: count$) of $\mathcal{X}$ where the word $w$ in the tuple is a seed word related to disease topic $k$, $t^{'}$ is equal to the time point $t$ in equation (\ref{eq:dirichlet_gamma}) and $l$ refers to any location in the set $L$. In other words, $N_{k,t}$ accounts for the occurrence of seed words related to topic $k$ in $\mathcal{X}$ at time point $t$. Higher occurrence of seed words indicates higher prominence of topic $k$ at time point $t$ and vice versa. Therefore, asymmetric prior $\gamma^{(k)}$ is used to incorporate prior information into the supervised topic model regarding prominence of disease topic $k$ at different time points. The hyperparameter $\gamma^{'}$ in equation (\ref{eq:dirichlet_gamma}) is an additional smoothing parameter that contributes a flat pseudocount to each component of $\gamma^{(k)}$. Additive smoothing is done to assign non-zero probabilities to those time points for which we have no prior information (zero occurrence of seed words) related to topic $k$.     

$\theta_{l}$ is also associated with an asymmetric Dirichlet prior parameterized by a $K$-dimensional vector $\alpha^{(l)}$ as defined below in equation (\ref{eq:dirichlet_alpha}).

\begin{equation}
  \label{eq:dirichlet_alpha}
  \alpha^{(l)} = [N_{l,1} + \alpha^{'}, N_{l,2} + \alpha^{'},\cdots, N_{l,k} + \alpha^{'},\cdots, N_{l,K} + \alpha^{'}]
\end{equation}

where, $N_{l,k}$ is the sum of the \textit{count} variable across those tuples ($\lbrace w, l^{'}, t\rbrace: count$) of $\mathcal{X}$ where the word $w$ in the tuple is a seed word related to disease topic $k$, $l^{'}$ is equal to the location $l$ in equation (\ref{eq:dirichlet_alpha}) and $t$ can be any time point in the range $\lbrace1,2,\cdots,T\rbrace$. In other words, $N_{l,k}$ accounts for the occurrence of seed words related to topic $k$ in $N_{l}$. The hyperparameter $\alpha^{'}$ is the additional smoothing parameter that contributes a non-zero pseudocount to each component of $\alpha^{(l)}$. Additive smoothing is done to assign non-zero probabilities to those locations for which we have no prior information (zero occurrence of seed words) related to topic $k$.  

Finally, seed topic distribution ($\phi^{s}$) and regular topic distribution ($\phi^{r}$) are drawn from  symmetric Dirichlet priors~\cite{wallach2009rethinking} where each component of the parameter vectors $\mu^{(k)}$  ($S$-dimensional) and $\beta^{(k)}$ ($V$-dimensional) assumes the values of the hyperparameters $\mu^{'}$ and $\beta^{'}$ respectively, i.e., $\mu^{(k)} = [\mu^{'}, \mu^{'}, \cdots, \mu^{'}]$ and  $\beta^{(k)} = [\beta^{'}, \beta^{'}, \cdots, \beta^{'}]$.

\paragraph*{Choice of hyperparameters.} A hyperparameter is defined as the parameter of a prior distribution. The hyperparameters $\alpha^{'}$, $\gamma^{'}$, $\beta^{'}$ and $\mu^{'}$ are set to $2/K$, $0.01$, $0.01$ and $1e-07$ respectively. These values are chosen heuristically, and an improved performance of the supervised topic model could be achieved via efficient hyperparameter optimization~\cite{wallach2009rethinking} . As suggested in Jagarlamudi et al.~\cite{jagarlamudi2012incorporating} , we set the sampling probability $\pi_{k}$ to a constant value of 0.7 for each topic $k \in \lbrace 1,2,\cdots,K\rbrace$.

\subsection*{Inference via collapsed gibbs sampling}

The key problem in the supervised topic model is posterior inference. This amounts to reversing the defined generative process and inferring the output (latent) parameters $\theta$, $\phi^r$ , $\phi^s$ and $\xi$ given the observed tuples in $N_{l}$. A standard approach of posterior inference in topic models is collapsed gibbs sampling~\cite{griffiths2004finding} , a Markov Chain Monte Carlo (MCMC) method. 

To estimate the model parameters $\theta$, $\phi^r$ , $\phi^s$ and $\xi$ via collapsed gibbs sampling, we need to compute the conditional probability distribution $\Pr(z_{i}=k|{\bf w},{\bf t}, {\bf l}, {\bf z}_{-i}, \alpha^{(l)}, \beta^{(k)}, \mu^{(k)}, \gamma^{(k)})$ where $z_{i}$ represents the topic assignment for the $i^{th}$ tuple or entry in $N_{l}$. ${\bf z}_{-i}$ represents the topic assignments for all entries in $N_{l}$ except the $i^{th}$ entry. We have three scenarios as shown below.

\begin{itemize}
\item If word $w_{i}$ in the $i^{th}$ entry of $N_{l}$ is a regular word and $k$ is a regular topic, then the conditional probability distribution is defined below in equation (\ref{eq:conditionalregularregular}).

\begin{align}
\LARGE
\label{eq:conditionalregularregular}
\Pr(z_{i}=k| {\bf w},{\bf t}, {\bf l}, {\bf z}_{-i},\alpha^{(l)},\beta^{(k)},\mu^{(k)},\gamma^{(k)}) &\propto \frac{n^{k,-i}_{w_{i}} + \beta^{'}}{\sum_{v = 1}^V (n^{k,-i}_{v} + \beta^{'})}  \cdot \frac{m^{k,-i}_{t_{i}} + \gamma^{(k)}_{t_{i}}}{\sum_{t = 1}^T (m^{k,-i}_{t} + \gamma^{(k)}_{t})}
\cdot  \frac{o^{l,-i}_{k} + \alpha^{(l)}_{k}}{\sum_{k^{'} = 1}^K (o^{l,-i}_{k^{'}} + \alpha^{(l)}_{k^{'}})} \nonumber \\
&\cdot \frac{(\sum_{v=1}^V n^{k,-i}_{v} + \beta^{'}) + \pi_k}{(\sum_{v=1}^V n^{k,-i}_{v} + \beta^{'}) + (\sum_{v=1}^S s^{k,-i}_{v} + \mu^{'}) + 2 \cdot \pi_k} 
\end{align}

\item If word $w_{i}$ in the $i^{th}$ entry of $N_{l}$ is a regular word and $k$ is a seed topic, then the conditional probability distribution $\Pr(z_{i}=k|{\bf w},{\bf t}, {\bf l}, {\bf z}_{-i}, \alpha^{(l)}, \beta^{(k)}, \mu^{(k)}, \gamma^{(k)}) = 0$ since a regular word cannot be generated from any of the seed topic distributions. 

\item If word $w_{i}$ in the $i^{th}$ entry of $N_{l}$ is a seed word, then word $w_{i}$ can be generated from either the seed topic $k$ or the regular topic $k$. If word $w_{i}$ is generated from a seed topic $k$, then the conditional probability distribution is defined below in equation (\ref{eq:conditionalseedseed}). On the other hand, if word $w_{i}$ is generated from a regular topic $k$, then the conditional probability distribution is defined below in equation (\ref{eq:conditionalseedregular}).

\begin{align}
\LARGE
\label{eq:conditionalseedseed}
\Pr(z_{i}=k| {\bf w},{\bf t}, {\bf l}, {\bf z}_{-i},\alpha^{(l)},\beta^{(k)},\mu^{(k)},\gamma^{(k)}) &\propto \frac{s^{k,-i}_{w_{i}} + \mu^{'}}{\sum_{v = 1}^S (s^{k,-i}_{v} + \mu^{'})}  \cdot \frac{m^{k,-i}_{t_{i}} + \gamma^{(k)}_{t_{i}}}{\sum_{t = 1}^T (m^{k,-i}_{t} + \gamma^{(k)}_{t})}
\cdot  \frac{o^{l,-i}_{k} + \alpha^{(l)}_{k}}{\sum_{k^{'} = 1}^K (o^{l,-i}_{k^{'}} + \alpha^{(l)}_{k^{'}})} \cdot  \pi_k
\end{align}

\begin{align}
\LARGE
\label{eq:conditionalseedregular}
\Pr(z_{i}=k| {\bf w},{\bf t}, {\bf l}, {\bf z}_{-i},\alpha^{(l)},\beta^{(k)},\mu^{(k)},\gamma^{(k)}) &\propto \frac{n^{k,-i}_{w_{i}} + \beta^{'}}{\sum_{v = 1}^V (n^{k,-i}_{v} + \beta^{'})}  \cdot \frac{m^{k,-i}_{t_{i}} + \gamma^{(k)}_{t_{i}}}{\sum_{t = 1}^T (m^{k,-i}_{t} + \gamma^{(k)}_{t})}
\cdot  \frac{o^{l,-i}_{k} + \alpha^{(l)}_{k}}{\sum_{k^{'} = 1}^K (o^{l,-i}_{k^{'}} + \alpha^{(l)}_{k^{'}})} \cdot  (1-\pi_k)
\end{align}

\end{itemize}

In equations (\ref{eq:conditionalregularregular}), (\ref{eq:conditionalseedseed}) and (\ref{eq:conditionalseedregular}), 
$n^{k,-i}_{w_{i}}$ denotes the number of times word $w_{i}$ is assigned to regular topic $k$ across all entries in $N_{l}$ except the $i^{th}$ 
entry, $s^{k,-i}_{w_{i}}$ denotes the number of times seed word $w_{i}$ is assigned to seed topic $k$ across all entries in $N_{l}$ except the $i^{th}$ entry, $m^{k,-i}_{t_{i}}$ denotes the number of times time point $t_{i}$ is assigned to topic $k$ across all entries in $N_{l}$ except the $i^{th}$ entry and 
$o^{l,-i}_{k}$ denotes the number of times location $l$ is associated with topic $k$ across all entries in $N_{l}$ except the $i^{th}$ entry. $\alpha^{(l)}_{k}$ refers to the $k^{th}$ component of $\alpha^{(l)}$ and $\gamma^{(k)}_{t_{i}}$ denotes the component of $\gamma^{(k)}$ corresponding to time point $t_{i}$.

\paragraph*{Implementing the collapsed gibbs sampler.} Collapsed gibbs sampler for the supervised topic model is surprisingly easy to implement. It involves setting up the required count variables, randomly initializing them, and then the gibbs sampler executes in an iterative fashion where on each iteration a topic is sampled for each entry in $N_{l}$ according to equation (\ref{eq:conditionalregularregular}) or equation (\ref{eq:conditionalseedseed}) and equation (\ref{eq:conditionalseedregular}) depending on whether the word in the entry is a regular word or a seed word respectively. The required count variables include $n^{k}_{w_{i}}$, $s^{k}_{w_{i}}$, $m^{k}_{t_{i}}$ and $o^{l}_{k}$ corresponding to the $i^{th}$ entry in $N_{l}$. For simplicity and efficiency, we also keep a running count of $n^{k}$ ($=\sum_{v=1}^{V} n^{k}_{v}$, the total number of times any word in vocabulary $V$ is assigned to topic $k$), $s^{k}$ ($=\sum_{v=1}^{S} s^{k}_{v}$, the total number of times any word in the set $S$ of seed words is assigned to the corresponding seed topic $k$), $m^{k}$ ($=\sum_{t=1}^{T} m^{k}_{t}$, the total number of times any time point $t \in \lbrace1,2,\cdots,K\rbrace$ is assigned to topic $k$) and $o^{l}$ ($=\sum_{k=1}^{K} o^{l}_{k}$, the total number of times any topic $k \in \lbrace1,2,\cdots,K\rbrace$ is associated with location $l$). Finally, in addition to the mentioned count variables, we also require an array ${\bf z}$ which will contain the topic assignment for each entry or tuple in $N_{l}$. Once we choose a topic for a particular entry in $N_{l}$, the chosen topic is set in the ${\bf z}$ array and the count variables are incremented in the appropriate position relevant to the entry. 
Following the gibbs iterations, the count variables can be used to compute the output (latent) parameters $\theta$, $\phi^r$ , $\phi^s$ and $\xi$ as shown below in equation (\ref{eq:updates}).

\begin{align}
\LARGE
\label{eq:updates}
\theta_{l,k} &= \frac{o^l_k + \alpha^{(l)}_k}{\sum_{k^{'}=1}^K (o^l_{k^{'}} + \alpha^{(l)}_{k^{'}})} \nonumber \\
\phi^{r}_{k,w} &= \frac{n^k_w + \beta^{'}}{\sum_{v=1}^V (n^k_v + \beta^{'})}  \nonumber \\
\phi^{s}_{k,w} &= \frac{s^k_w + \mu^{'}}{\sum_{v=1}^S (s^k_v + \mu^{'})}  \\
\xi_{k,t} &= \frac{m^k_t + \gamma^{(k)}_t}{\sum_{t=1}^T (m^k_t + \gamma^{(k)}_t)} \nonumber
\end{align}

where, $\theta_{l,k}$ represents the probability of topic $k$ given location $l$, $\phi^{r}_{k,w}$ represents the
probability of word $w$ given topic $k$, $\phi^{s}_{k,w}$ represents the probability of seed word $w$ given seed 
topic $k$ and $\xi_{k,t}$ denotes the temporal trend value of topic $k$ at time point $t$. We ran the gibbs sampler for 300 iterations.

\subsection*{Baseline methods for case count estimation} 

We compared {\fullmodelforecast} with 2 baseline methods, namely {\baselinemodel} and {\fullmodelbaseline}. In {\baselinemodel}, we fitted an autoregressive-moving-average model (ARMA($p$, $q$)~\cite{box2011time}) over past disease case counts to generate case count estimates as shown below in equation (\ref{eq:ARMA}). 

\begin{equation}
\label{eq:ARMA}
\hat{S}_{D,T} = \epsilon_{D,T} + {\sum_{i=1}^p{\gamma_iS_{D,T-i}}} + {\sum_{i=1}^q{\theta_i\epsilon_{D,T-i}}} 
\end{equation}

where, $p$ and $q$ are the orders of the autoregressive (AR) and moving 
average (MA) components, respectively. $\epsilon_{D, T}, \epsilon_{D, T-1}, ...., \epsilon_{D, T-q}$ represent the white noise error terms. For further details including boundary conditions of ARMA, please refer to Box et al.~\cite{box2011time}. {\baselinemodel} doesn't use any information related to temporal topic trends ($\xi_{z}$). However, in {\fullmodelbaseline}, we used an autoregressive–moving-average model with external input variables (ARMAX($p$,$q$)~\cite{box2011time}). As shown below in equation (\ref{eq:ARMAX}), ARMAX($p$, $q$) incorporates information from both past case counts and temporal topic trends ($\xi_{z}$) in order to estimate case counts. Similar to {\fullmodelforecast}, external input variables are represented by the temporal topic trends ($\xi_{z}$).   

\begin{equation}
\label{eq:ARMAX}
\hat{S}_{D,T} = \epsilon_{D,T} + {\sum_{i=1}^p{\gamma_iS_{D,T-i}}} + {\sum_{i=0}^p{\eta_i\xi_{z,T-i}}} + {\sum_{i=1}^q{\theta_i\epsilon_{D,T-i}}} 
\end{equation}

where, $p$ and $q$ are the orders of the autoregressive (AR) and moving average (MA) components, respectively. For further details, please refer to Box et al.~\cite{box2011time}.

\begin{figure*}[ht]
\centering
\includegraphics[width=\linewidth]{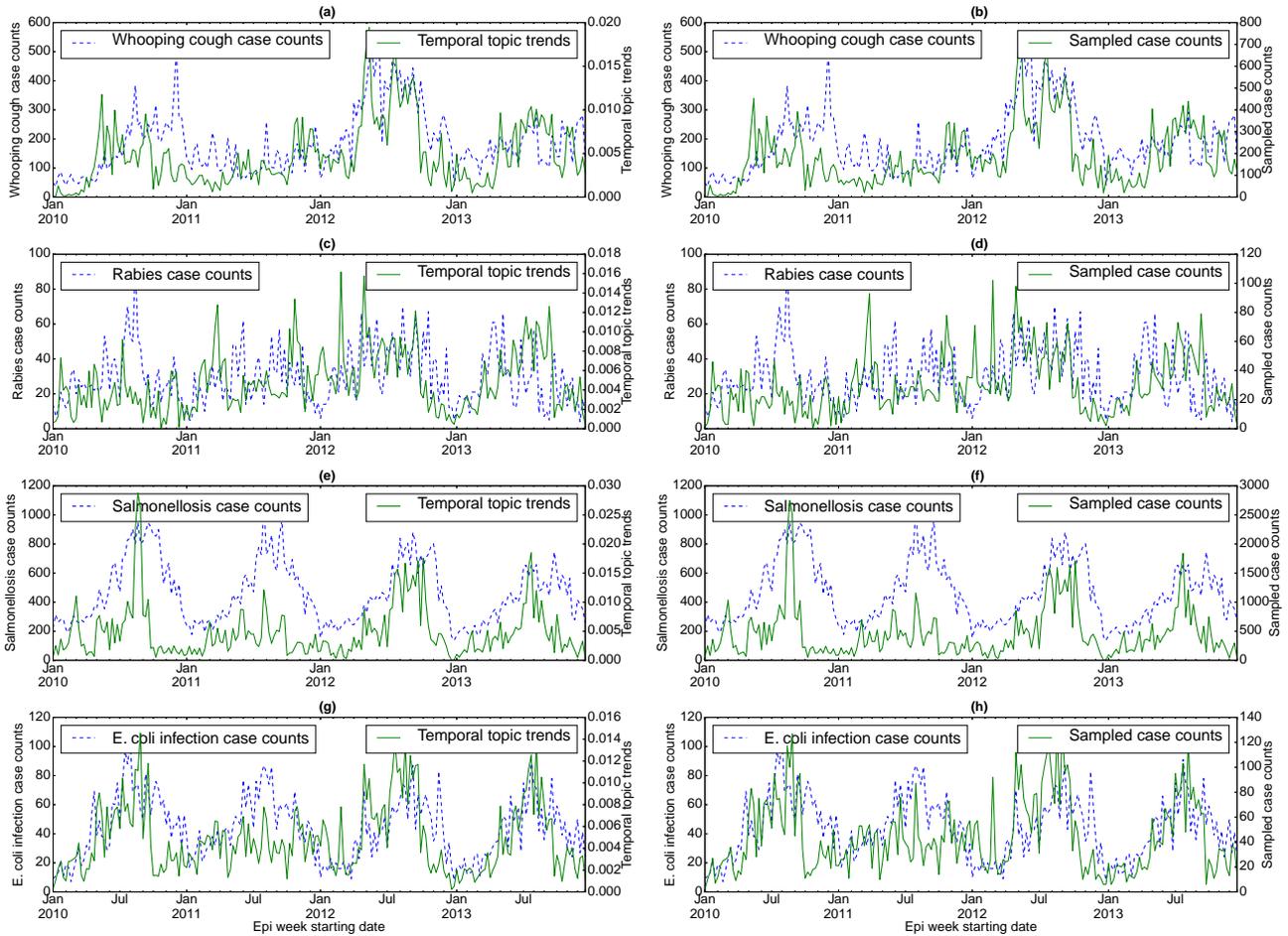}
\caption{\textbf{Correlation between disease case counts and temporal topic distributions or trends ($\xi_{z}$) extracted by {\fullmodel} for (a) whooping cough, (c) rabies, (e) salmonellosis, and (g) E. coli infection in U.S.} Along with the temporal topic trends ($\xi_{z}$), we also showed the correlation between disease case counts and sampled case counts (generated by multinomial sampling from temporal topic trends) for (b) whooping cough, (d) rabies, (f) salmonellosis, and (h) E. coli infection. Note, the sampled case counts and disease case counts share almost similar numerical range. However, the temporal topic trend values are at different numerical range (ranging from 0 to 1) with respect to the disease case counts.}
\label{fig:US_disease_news}
\end{figure*}

\begin{figure*}[ht]
\centering
\includegraphics[width=\linewidth]{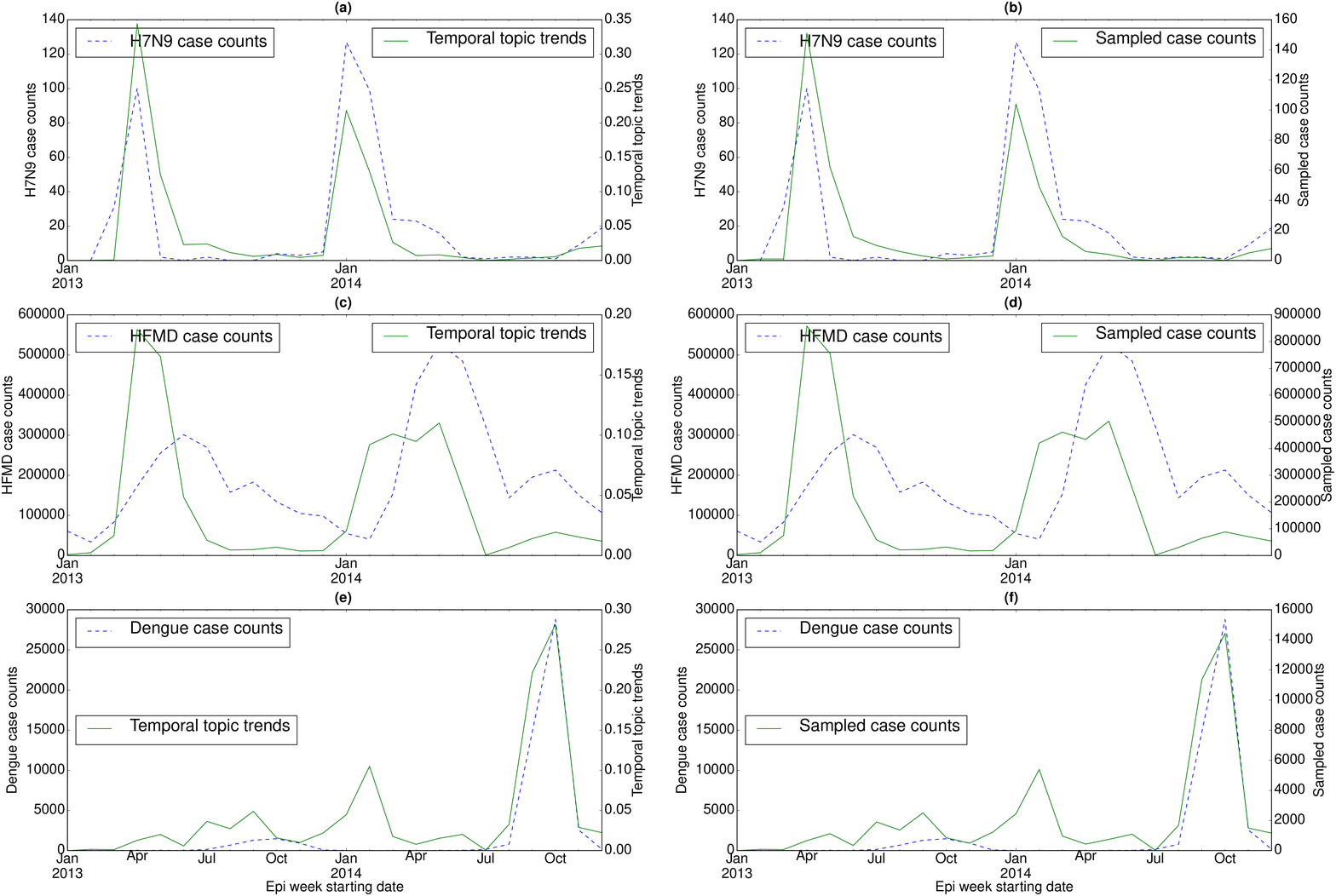}
\caption{\textbf{Correlation between disease case counts and temporal topic distributions or trends ($\xi_{z}$) extracted by {\fullmodel} for (a) H7N9, (c) HFMD, and (e) dengue in China.} Along with the temporal topic trends ($\xi_{z}$), we also showed the correlation between disease case counts and sampled case counts (generated by multinomial sampling from temporal topic trends) for (b) H7N9, (d) HFMD, and (f) dengue. Note, the sampled case counts and disease case counts share almost similar numerical range. However, the temporal topic trend values are at different numerical range (ranging from 0 to 1) with respect to the disease case counts.}
\label{fig:China_disease_news}
\end{figure*}

\begin{figure*}[ht]
\centering
\includegraphics[width=\linewidth]{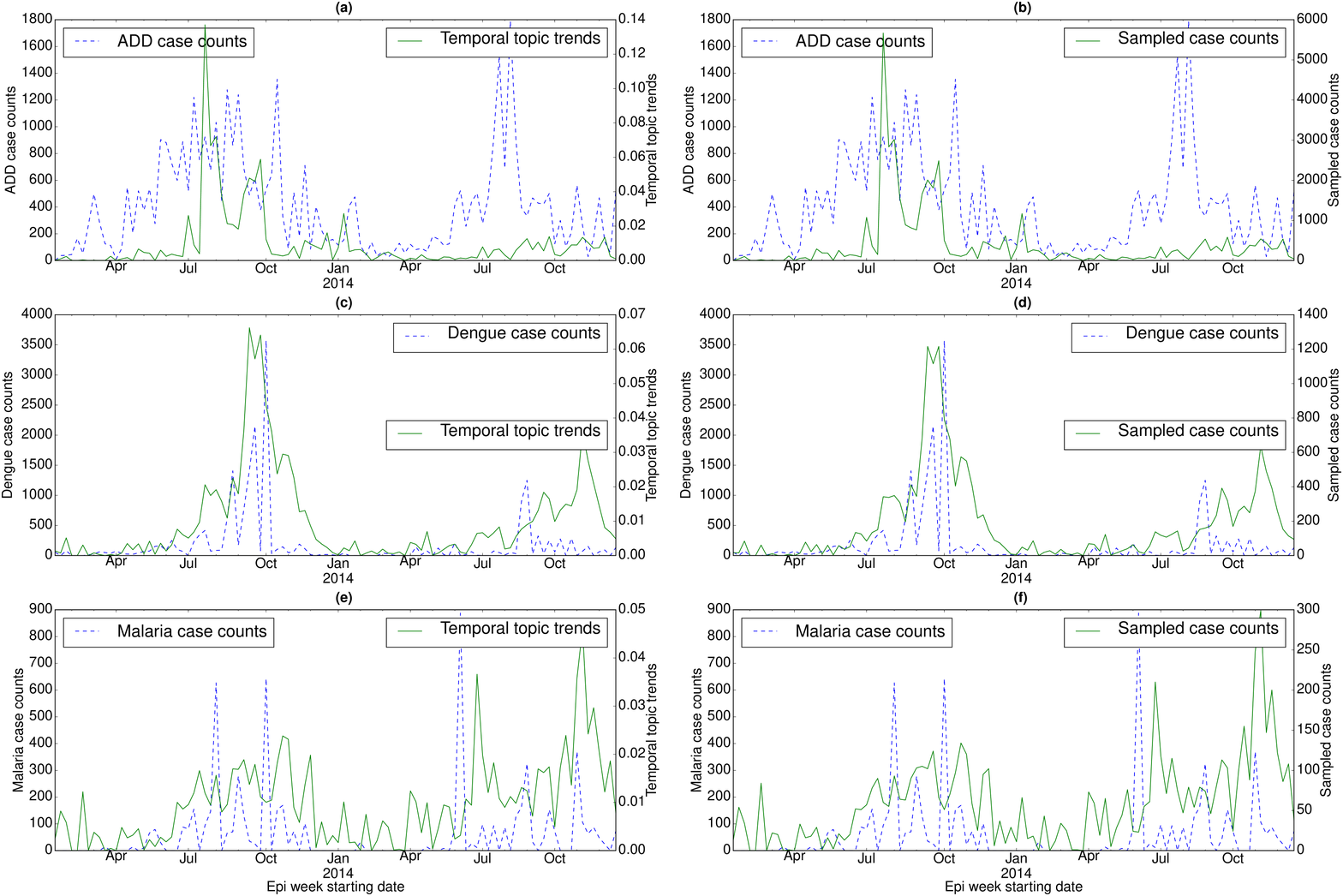}
\caption{\textbf{Correlation between disease case counts and temporal topic distributions or trends ($\xi_{z}$) extracted by {\fullmodel} for (a) ADD, (c) dengue, and (e) malaria in India.} Along with the temporal topic trends ($\xi_{z}$), we also showed the correlation between disease case counts and sampled case counts (generated by multinomial sampling from temporal topic trends) for (b) ADD, (d) dengue, and (f) malaria. Note, the sampled case counts and disease case counts share almost similar numerical range. However, the temporal topic trend values are at different numerical range (ranging from 0 to 1) with respect to the disease case counts.}
\label{fig:India_disease_news}
\end{figure*}  

\begin{figure*}[ht]
\centering
\includegraphics[width=\linewidth]{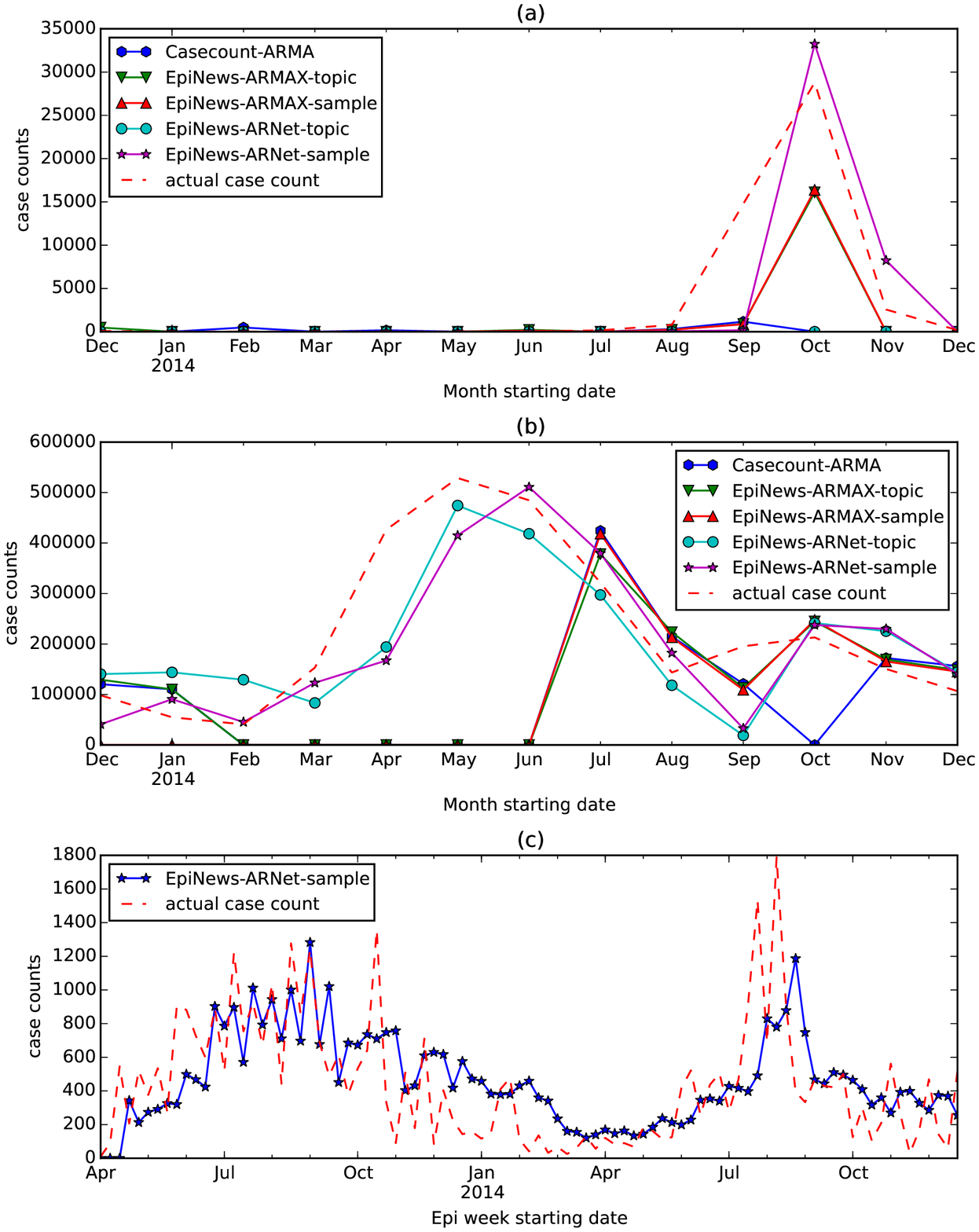}
\caption{\textbf{Temporal correlation between actual case counts and case counts estimated by the methods {\baselinemodel}, {\fullmodelbaseline} and {\fullmodelforecast} corresponding to (a) dengue and (b) HFMD in China.} In (a) and (b), {\fullmodelbaseline}-topic and {\fullmodelforecast}-topic use temporal topic trends as external variables. On the other hand, {\fullmodelbaseline}-sample and {\fullmodelforecast}-sample use sampled case counts as external variables. In (c), we showed the temporal correlation between actual case counts and case counts estimated by {\fullmodelforecast}-sample corresponding to ADD in India.}
\label{fig:China_disease_estimate}
\end{figure*}

\begin{table*}[ht!]
\centering
\tiny
\resizebox{\columnwidth}{!}{%
\begin{tabular}{|l||l||l||l|}
\hline
{Whooping cough topic} & {Rabies topic} &
{Salmonellosis topic}  & {E. coli infection topic} \\
\hline
\hline
{Seed words} & {Seed words} & {Seed words} & {Seed words} \\
\hline
\begin{tabular}[l]{@{}l|l@{}}
child  &  0.1498  \\
school  &  0.1068  \\
cough  &  0.0828  \\
pertussis  &  0.0701  \\
whoop  &  0.0691  \\
whooping  &  0.0679  \\
infant  &  0.0596  \\
student  &  0.0557  \\
contagious  &  0.0454  \\
booster  &  0.0406  \\
cold  &  0.0395  \\
coughing  &  0.0309  \\
nose  &  0.0304  \\
respiratory  &  0.0284  \\
mild  &  0.0269  \\
tdap  &  0.0231  \\
immunize  &  0.0212  \\
runny  &  0.0198  \\
tetanus  &  0.0175  \\
breathe  &  0.0144 
\end{tabular} 
& 
\begin{tabular}[l]{@{}l|l@{}}
animal  &  0.1596  \\
rabies  &  0.1191  \\
rabid  &  0.0718  \\
bite  &  0.0695  \\
rabie  &  0.0674  \\
virus  &  0.0649  \\
wild  &  0.0585  \\
bat  &  0.0472  \\
raccoon  &  0.0471  \\
skunk  &  0.0424  \\
fox  &  0.0422  \\
wildlife  &  0.0379  \\
domestic  &  0.0323  \\
saliva  &  0.0247  \\
scratch  &  0.0237  \\
quarantine  &  0.0213  \\
horse  &  0.0192  \\
viral  &  0.0190  \\
livestock  &  0.0166  \\
mammal  &  0.0156  
\end{tabular}
&
\begin{tabular}[l]{@{}l|l@{}}
food  &  0.2056  \\
salmonella  &  0.1031  \\
product  &  0.1013  \\
recall  &  0.0878  \\
drug  &  0.0712  \\
consumer  &  0.0705  \\
contamination  &  0.0598  \\
fda  &  0.0579  \\
contaminate  &  0.0567  \\
abdominal  &  0.0351  \\
egg  &  0.0277  \\
chicken  &  0.0275  \\
poultry  &  0.025  \\
arthritis  &  0.0145  \\
peanut  &  0.0139  \\
cantaloupe  &  0.01  \\
shell  &  0.0086  \\
typhimurium  &  0.0083  \\
newport  &  0.0082  \\
enteritidis  &  0.0074  
\end{tabular}
&
\begin{tabular}[l]{@{}l|l@{}}
coli  &  0.2265  \\
boil  &  0.0887  \\
cell  &  0.0745  \\
toxin  &  0.0628  \\
escherichia  &  0.0617  \\
clinical  &  0.0573  \\
chemical  &  0.0557  \\
kidney  &  0.0414  \\
microbiology  &  0.0402  \\
reaction  &  0.0397  \\
hemolytic  &  0.0376  \\
lettuce  &  0.0366  \\
uremic  &  0.036  \\
physical  &  0.0342  \\
gene  &  0.0339  \\
shiga  &  0.0202  \\
expression  &  0.0162  \\
chemistry  &  0.0149  \\
stec  &  0.0125  \\
biochemistry  &  0.0094 
\end{tabular} \\
\hline \hline
{\makecell{Regular words \\ with higher probabilities}} & {\makecell{Regular words \\ with higher probabilities}} & {\makecell{Regular words \\ with higher probabilities}} & {\makecell{Regular words \\ with higher probabilities}} \\ 
\hline
\begin{tabular}[l]{@{}l|l@{}}
contact  &  0.0037  \\
young  &  0.0023  \\
adult  &  0.0022  \\
vaccination  &  0.0019  \\
vaccine  &  0.0019  \\
california  &  0.0019  \\
vaccinate & 0.0018 \\
parent  &  0.0017  \\
woman & 0.0015 \\
baby  &  0.0014  \\
immunization  &  0.0011  \\
kid  &  0.0009  \\
air  &  0.0008  \\
weather  &  0.0007  \\
pregnant  &  0.0006  \\
mother  &  0.0006  \\
dose  &  0.0006  \\
antibiotic  &  0.0005  \\
pneumonia  &  0.0003  
\end{tabular} 
&
\begin{tabular}[l]{@{}l|l@{}}
pet  &  0.0037  \\
contact  &  0.0037  \\
cat  &  0.0028  \\
vaccination  &  0.0024  \\
florida  &  0.0015  \\
vaccine  &  0.0014  \\
shot  &  0.0014  \\
street  &  0.0013  \\
clinic  &  0.0012  \\
texas  &  0.0010  \\
park  &  0.0010  \\
york  &  0.0010  \\
wound  &  0.0009  \\
virginia  &  0.0008  \\
ferret  &  0.0007  \\
brain  &  0.0007  \\
coyote  &  0.0005  \\
nervous  &  0.0005  \\
canine  &  0.0002  
\end{tabular} 
&
\begin{tabular}[l]{@{}l|l@{}}
eat  &  0.0019  \\
diarrhea  &  0.0019  \\
nausea  &  0.0013  \\
foodborne  &  0.0013  \\
package  &  0.0012  \\
contaminated  &  0.0011  \\
meat  &  0.0011  \\
restaurant  &  0.0010  \\
vomit  &  0.0010  \\
products  &  0.0008  \\
cook  &  0.0008  \\
beef  &  0.0008  \\
raw  &  0.0007  \\
temperature  &  0.0006  \\
honey  &  0.0005  \\
pepper  &  0.0004  \\
weather  &  0.0003  \\
salad  &  0.0003  \\
mango  &  0.0002  
\end{tabular} 
& 
\begin{tabular}[l]{@{}l|l@{}}
transmit  &  0.0014  \\
massachusetts  &  0.0013  \\
surface & 0.0012 \\
body & 0.0012 \\
pennsylvania  &  0.0012  \\
blood  &  0.0012  \\
pathogen & 0.0011 \\
resistant & 0.0011 \\
drinking & 0.0011 \\
agricultural & 0.0011 \\
hygiene & 0.0010 \\
raw  &  0.0009  \\
apple & 0.0009 \\
sandwich & 0.0009 \\
milk  &  0.0008  \\
stool  &  0.0008  \\
parasite  &  0.0005  \\
acs  &  0.0002  \\
receptor  &  0.0001  
\end{tabular}
\\
\hline
\end{tabular}%
}
\caption{\label{topicwordtable_US} \textbf{Four disease topics (Whooping Cough, Rabies, Salmonella and E. coli infection) discovered by the supervised topic model from the HealthMap corpus for U.S.} For each disease topic, we show the seed words and their corresponding probabilities in the seed topic distribution. Along with the seed words, we also show some of the regular words (having higher probabilities in the regular topic distribution) discovered by the supervised topic model related to these input seed words.}
\end{table*}
\begin{table*}[ht!]
\centering
\tiny
\resizebox{\columnwidth}{!}{%
\begin{tabular}{|l||l||l||l|}
\hline
{H7N9 topic} & {HFMD topic} &
{Dengue topic} \\
\hline
\hline
{Seed words} & {Seed words} & {Seed words} \\
\hline
\begin{tabular}[l]{@{}l|l@{}}
flu & 0.1229 \\
bird & 0.1225 \\
avian & 0.1053 \\
influenza & 0.1051 \\
human & 0.1031 \\
virus & 0.0832 \\
poultry & 0.0786 \\
market & 0.0610 \\
animal & 0.0360 \\
chicken & 0.0303 \\
respiratory & 0.0230 \\
spring & 0.0227 \\
farm & 0.0224 \\
farmer & 0.0213 \\
slaughter & 0.0194 \\
winter & 0.0179 \\
egg & 0.0125 \\
pandemic & 0.0117 \\
h7n9 & 0.0012 \\
h5n1 & 0.0000 

\end{tabular} 
& 
\begin{tabular}[l]{@{}l|l@{}}
hand & 0.1573 \\
child & 0.1384 \\
mouth & 0.1127 \\
school & 0.1016 \\
foot & 0.0916 \\
class & 0.0734 \\
hfmd & 0.0557 \\
parent & 0.0546 \\
nursery & 0.0343 \\
kindergarten & 0.0294 \\
oral & 0.0192 \\
intestinal & 0.0185 \\
infant & 0.0178 \\
mumps & 0.0174 \\
measles & 0.0172 \\
herpes & 0.0140 \\
enterovirus & 0.0135 \\
encephalitis & 0.0124 \\
dysentery & 0.0117 \\
ulcer & 0.0093
\end{tabular}
&
\begin{tabular}[l]{@{}l|l@{}}
fever & 0.2269 \\
dengue & 0.1586 \\
mosquito & 0.1052 \\
october & 0.0826 \\
water & 0.0682 \\
breeding & 0.0559 \\
street & 0.0481 \\
bite & 0.0330 \\
aedes & 0.0317 \\
pain & 0.0294 \\
breed & 0.0280 \\
park & 0.0269 \\
sanitation & 0.0179 \\
borne & 0.0175 \\
albopictus & 0.0168 \\
rain & 0.0139 \\
hemorrhagic & 0.0125 \\
vector & 0.0115 \\
larva & 0.0089 \\
aegypti & 0.0066 
\end{tabular} \\
\hline \hline
{\makecell{Regular words \\ with higher probabilities}} & {\makecell{Regular words \\ with higher probabilities}} & {\makecell{Regular words \\ with higher probabilities}} \\ 
\hline
\begin{tabular}[l]{@{}l|l@{}}
zhejiang & 0.0034 \\ 
beijing & 0.0034 \\
shanghai & 0.0030 \\
agriculture & 0.0015 \\
pneumonia & 0.0013 \\
temperature & 0.0011 \\
food & 0.0010 \\
eat & 0.0009 \\
duck & 0.0008 \\
pigeon & 0.0008 \\
cook & 0.0006 \\
vaccine & 0.0006 \\
tamiflu & 0.0005 \\
meat & 0.0004 \\
strain & 0.0004 \\
raw & 0.0003 \\
pig & 0.0003 
\end{tabular} 
&
\begin{tabular}[l]{@{}l|l@{}}
shandong & 0.0028 \\
hunan & 0.0025 \\
care & 0.0015 \\
rash & 0.0008 \\
meningitis & 0.0007 \\
viral & 0.0007 \\
hepatitis & 0.0007 \\
body & 0.0006 \\
tuberculosis & 0.0006 \\
childhood & 0.0005 \\
palm & 0.0004 \\
organ & 0.0003 \\
skin & 0.0003 \\
buttock & 0.0003 \\
childcare & 0.0003 \\
blister & 0.0002 \\
kidney & 0.0002 
\end{tabular} 
& \begin{tabular}[l]{@{}l|l@{}}
guangdong & 0.0071 \\
guangzhou & 0.0056 \\
site & 0.0013 \\
temperature & 0.0010 \\
weather & 0.0009 \\
muscle & 0.0008 \\
blood & 0.0006 \\
urban & 0.0005 \\
bleed & 0.0004 \\
diarrhea & 0.0004 \\
medicine & 0.0004 \\
stagnant & 0.0004 \\
spray & 0.0003 \\
rainy & 0.0003 \\
climate & 0.0003 \\
cough & 0.0002 \\
tank & 0.0002 
\end{tabular}
\\
\hline
\end{tabular}%
}
\caption{\label{topicwordtable_China} \textbf{Three disease topics (H7N9, HFMD and dengue) discovered by the supervised topic model from the HealthMap corpus for China.} For each disease topic, we show the seed words and their corresponding probabilities in the seed topic distribution. Along with the seed words, we also show some of the regular words (having higher probabilities in the regular topic distribution) discovered by the supervised topic model related to these input seed words.}
\end{table*}
\begin{table*}[ht!]
\centering
\tiny
\resizebox{\columnwidth}{!}{%
\begin{tabular}{|l||l||l||l|}
\hline
{ADD topic} & {Dengue topic}&
{Malaria topic} \\
\hline
\hline
{Seed words} & {Seed words} & {Seed words} \\
\hline
\begin{tabular}[l]{@{}l|l@{}}
fall & 0.1284 \\
child & 0.1148 \\
school & 0.0949 \\
student & 0.0868 \\
food & 0.0837 \\
consume & 0.0611 \\
eat & 0.0588 \\
vomit & 0.0549 \\
meal & 0.0525 \\
stomach & 0.0412 \\
diarrhea & 0.0315 \\
nausea & 0.0304 \\
vomiting & 0.0300 \\
poisoning & 0.0249 \\
poison & 0.0241 \\
midday & 0.0237 \\
contaminated & 0.0183 \\
cook & 0.0179 \\
lunch & 0.0117 \\
contaminate & 0.0105  
\end{tabular} 
& 
\begin{tabular}[l]{@{}l|l@{}}
dengue & 0.2090 \\
fever & 0.0978 \\
municipal & 0.0759 \\
breeding & 0.0658 \\
borne & 0.0586 \\
mosquito & 0.0555 \\
september & 0.0491 \\
august & 0.0429 \\
water & 0.0408 \\
rain & 0.0385 \\
aedes & 0.0382 \\
ward & 0.0382 \\
platelet & 0.0330 \\
breed & 0.0300 \\
larva & 0.0268 \\
blood & 0.0264 \\
bite & 0.0246 \\
chikungunya & 0.0206 \\
vector & 0.0199 \\
monsoon & 0.0084 
\end{tabular}
&
\begin{tabular}[l]{@{}l|l@{}}

malaria & 0.1504 \\
mosquito & 0.1166 \\
site & 0.0994 \\
water & 0.0893 \\
awareness & 0.0826 \\
lead & 0.0735 \\
vector & 0.0678 \\
breed & 0.0567 \\
monsoon & 0.0484 \\
blood & 0.0414 \\
construction & 0.0331 \\
camp & 0.0316 \\
drug & 0.0228 \\
rainfall & 0.0175 \\
typhoid & 0.0148 \\
tribal & 0.0133 \\
falciparum & 0.0114 \\
economic & 0.0110 \\
anopheles & 0.0099 \\
plasmodium & 0.0084 
\end{tabular} \\ 
\hline \hline
{\makecell{Regular words \\ with higher probabilities}} & {\makecell{Regular words \\ with higher probabilities}} & {\makecell{Regular words \\ with higher probabilities}} \\ 
\hline
\begin{tabular}[l]{@{}l|l@{}}
village & 0.0032 \\  
bihar & 0.0023 \\
inflammatory & 0.0020 \\
sample & 0.0018 \\
odisha & 0.0017 \\
ache & 0.0011 \\                                                       sick & 0.0010 \\
pain & 0.0008 \\                                                        iron & 0.0008 \\ 
rice & 0.0006 \\                                                   pesticide & 0.0005 \\                                                          flood & 0.0004 \\                                                          drink & 0.0004 \\                            
sanitation & 0.0004 \\                                                         stale & 0.0003 \\                                                       drinking & 0.0001 \\  
\end{tabular} 
&
\begin{tabular}[l]{@{}l|l@{}}
civic & 0.0038 \\  
delhi & 0.0026 \\
virus & 0.0018 \\                                                        temperature & 0.0014 \\
fogging & 0.0014 \\
haryana & 0.0009 \\
spray & 0.0009 \\
stagnant & 0.0008 \\                                                    infection & 0.0008 \\
aegypti & 0.0008 \\
drain & 0.0007 \\
larval & 0.0006 \\
stagnate & 0.0005 \\
gutter & 0.0003 \\                                                      rainwater & 0.0002 \\                                                    urbanization & 0.0001   
\end{tabular} 
& \begin{tabular}[l]{@{}l|l@{}}
mumbai & 0.0025 \\
virus & 0.0015 \\
maharashtra & 0.0011 \\
stagnant & 0.0011 \\ 
insect & 0.0009 \\
garbage & 0.0008 \\           
flu & 0.0008 \\                                               
spraying & 0.0007 \\                                          
aegypti & 0.0007 \\                                            
parasite & 0.0006 \\                                          
tank & 0.0006 \\                                                        
leptospirosis & 0.0005 \\                                      
urban & 0.0004 \\
drainage & 0.0003 \\
rainwater & 0.0002 \\
waterlog & 0.0002
\end{tabular}
\\ 
\hline
\end{tabular}%
}
\caption{\label{topicwordtable_India} \textbf{Three disease topics (ADD, dengue and malaria) discovered by the supervised topic model from the HealthMap corpus for India.} For each disease topic, we show the seed words and their corresponding probabilities in the seed topic distribution. Along with the seed words, we also show some of the regular words (having higher probabilities in the regular topic distribution) discovered by the supervised topic model related to these input seed words.}
\end{table*}

\end{document}